\documentclass[aps,prc,superscriptaddress,preprintnumbers,footinbib]{revtex4}
\usepackage{graphicx}
\newcommand{\be}{\begin{equation}}
\newcommand{\ee}{\end{equation}}

\begin{document}

\title{Progress in Understanding the Nuclear Equation of State 
at the Quark Level}

\author{A.W. Thomas$^{1,2}$ and P.A.M. Guichon$^3$\\
$^1$Thomas Jefferson National Accelerator Facility,\\ 
12000 Jefferson Ave., Newport News, VA 23606, USA\\
$^2$College of William and Mary, Williamsburg, VA 23187, USA \\
$^3$SPhN-DAPNIA, CEA Saclay, F91191 Gif sur Yvette, France\\
}

\begin{abstract}
At the present time there is a lively debate within the nuclear community concerning
the relevance of quark degrees of freedom in understanding nuclear
structure. We outline the key issues and review the
impressive progress made recently within the framework of the quark-meson
coupling model. In particular, we explain in quite general terms how
the modification of the internal structure of hadrons in-medium leads
naturally to three- and four-body forces, or equivalently, to density
dependent effective interactions.
\end{abstract}
\maketitle

\section{Introduction}

We have a fundamental theory of the strong interaction, namely 
Quantum Chromodynamics (QCD),
the mathematical beauty of which, in combination with its phenomenological
successes, has convinced most physicists that it must be correct.
Its predictions have been accurately confirmed in the short-distance
(or high energy) regime of {}``asymptotic freedom''. 
In the opposite limit of long
distances (corresponding to quark confinement), it will be extensively 
tested over the next decade through advances in lattice QCD combined
with new experimental facilities, such as the 12 GeV upgrade at Jefferson
Lab and the new hadronic capabilities at J-PARC and FAIR. One of the
most compelling challenges in modern nuclear physics is to relate the
properties of nuclear matter, from finite nuclei to neutron stars,
to the underlying quark and gluon structure of matter and QCD itself.

Over the past 50 years the standard approach to understanding nuclear
structure has involved the application of non-relativistic many-body theory,
based upon nucleon-nucleon (NN) potentials fit to experimental two-nucleon
scattering data. Amongst the potentials used at various stages we
mention the Paris potential~\cite{Cottingham:1973wt}, with its intermediate
range structure determined through dispersion relations and hence
almost model independent. Other approaches which are still widely
used are the one-boson-exchange forces, 
such as the various Bonn~\cite{Machleidt:1987hj}
and Nijmegen~\cite{deSwart:1995mb} potentials. In its modern form
this approach is best represented by the Argonne 18 potential~\cite{Wiringa:1994wb},
supplemented by a phenomenological three-body force~\cite{Carlson:1983kq}.
This has been used in combination with Green function monte-carlo
methods to calculate the energy levels of light nuclei.

These models of the nuclear forces have no direct connection to QCD, 
in that they
operate at the purely hadronic level. Another recent approach, which
preserves the chiral symmetry of QCD~\cite{Epelbaum:2004fk}, is
based on effective field theory (EFT). Such calculations are just
now achieving the phenomenological success in fitting NN data that
is characteristic of the Bonn and Nijmegen potentials mentioned above
-- with a similar number of fitting parameters, 
typically between 20 and 30.
The apparatus of EFT is also being exploited to generate the 
corresponding three-body
forces~\cite{Epelbaum:2005pn}, so that one can tackle the energy
levels of light nuclei. While the use of EFT permits one to preserve
the chiral symmetry of QCD, this is a pale reflection of the full
power of QCD itself. Indeed, given the extensive knowledge of chiral
symmetry at the time, one could have carried out the EFT program for the 
NN force in the 1960s, before the discovery of QCD.

Neither the usual treatments of nuclear structure outlined above, 
nor the usual EFT approach, have yet incorporated the requirements of special 
relativity. We mentioned earlier the results of the
Paris group's dispersion relation analysis of the nature of the intermediate
range NN force. The intermediate range NN force is unambiguously dominated
by two-pion exchange with a Lorentz scalar and isoscalar character.
Quantum Hadrodynamics (QHD) exploited this by formulating a description
of nuclear matter in which Lorentz scalar attraction (represented
by a $\sigma$ meson) competed with Lorentz vector repulsion (represented
by an $\omega$ meson)~\cite{Serot:1984ey}. At the saturation density
of nuclear matter, $\rho_{0}$, the mean scalar field strength in
QHD was of order 400 MeV. A similar result was found in a relativistic
Brueckner Hartree-Fock calculation using a boson 
exchange potential~\cite{Brockmann:1996xy}.
Given such a huge scalar field, with a strength almost half of the
mass of the nucleon itself, one must anticipate that it will have
a significant effect on the internal structure of the nucleons making
up the nuclear matter. In particular, we recall that the typical energy
associated with the excitation of internal degrees of freedom in the
nucleon is only 300 MeV (for spin excitations) to 500 MeV (for orbital
excitations). 

Let us briefly summarize. Any serious treatment of the model independent
Lorentz scalar nature of the intermediate range NN force leads to
the conclusion that the bound nucleons experience a scalar potential
of a strength up to one half of the mass of the nucleon itself --
a strength comparable with the typical internal excitation energy
of the nucleon. This immediately suggests that the
internal structure of the nucleon should indeed play a crucial role in
nuclear structure. The popular prejudice which leads many to neglect this
role is generally based on the observation that the nucleon binding energy
is very small in comparison with its mass. However this small binding
actually arises from a cancellation between the scalar and the vector
potential in-medium, while the structure modification is 
expected to arise primarily through the scalar
potential.

In order to anticipate the possible consequences of this 
very general observation, we
turn to atomic and molecular systems where we have text book experience.
We know that when an atom is subjected to a strong electric field,
its electron structure will rearrange in order to oppose the applied
field. This change in the internal structure of the atom, at least
if one is concerned solely with describing the energy of the system,
can be described in terms of an electric polarizability. In particular,
the energy of the system has a term quadratic in the applied electric
field, with the coefficient being (one half of) 
the electric polarizability. Exactly
the same thing happens if we apply a magnetic field, with the coefficient
of the term quadratic in the applied magnetic field involving the magnetic
polarizability.

Turning to the nucleon itself, we know that applied electric and magnetic
fields alter its internal structure, giving rise to electric and magnetic
polarizabilities and even, in sophisticated electron scattering experiments,
the so-called generalized polarizabilities~\cite{Guichon:1995pu,Thomas_Weise}.
Given this background it is remarkable that it has taken so long for
nuclear physicists to pay serious attention to the response of the
nucleon to an applied scalar field, especially its scalar polarizability.

Taking the experience with atomic physics as our lead, we would naturally
expect that a nucleon embedded in matter would have an energy with
a non-linear dependence on the mean scalar field: \be
M_N^{*} = M_N - g_\sigma \sigma + \frac{d}{2} (g_\sigma \sigma)^2
\label{eq:defn} \ee
where $d$ is the scalar polarizability. In QHD such a term is, of
course, absent -- although later phenomenological developments involving
a self-interaction of the scalar field can be rewritten (using a redefinition
of the scalar field) in just this way. Viewed in this way the scalar
polarizability can be seen as giving a very natural explanation of
such a self-interaction. In the first examination of nuclear matter
from this point of view, using the MIT bag model to describe the quark
structure of the bound nucleons, Guichon found exactly such a behavior,
with $d = +0.22R$ and $R$ the bag radius~\cite{Guichon:1987jp}. Since
that time, this model, which is known as the quark-meson coupling
(QMC) model, has been extended to describe finite nuclei~\cite{Guichon:1995ue}.
It has also been widely applied to the extremely interesting problem
of how the structure of hadrons in general might be expected to change
in-medium~\cite{Saito:2005rv,Saito:1996yb} -- applications which have included
the possibility of $\eta$- or $\omega$-nucleus bound 
states~\cite{Tsushima:1998qw}, the
structure of hypernuclei~\cite{Tsushima:1997cu} and the 
modification of the electric and
magnetic form factors of bound protons~\cite{Lu:1997mu}, 
as well as their structure functions~\cite{Saito:1993yw}.

Rather than reviewing these many applications, which have recently
been the subject of a major review~\cite{Saito:2005rv}, we return
to the atomic and molecular analogy for further insights which, as
we shall see, are crucial to understanding the structure of atomic
nuclei. When two atoms approach each other they mutually
modify their internal structure. This implies that when a third atom
approaches it does not see a pair of bare atoms. Consequently, the
total energy of the system will not be the sum of the individual pair-wise
interactions (potentials) -- that is: \be
V_{\textrm{tot}} \neq V_{12} + V_{23} + V_{13} \label{ew:3body} \ee
and the difference is, by definition, a {}``three-body force'',
$V_{123}$. By analogy, we anticipate that the scalar polarizability
of the nucleon will naturally lead to many-body forces in-medium.

In the next section we briefly review the exploration of the origin
of many-body forces (or equivalently density dependent effective interactions),
arising as a consequence of the modification of the internal structure
of the nucleon in-medium, within the framework of QMC. We shall see that
it is indeed possible, starting from the quark level, to generate
realistic effective interactions of the Skyrme type~\cite{Vautherin:1971aw}.

\section{Effective Forces of the Skyrme Type Derived from the Quark Level}

In 2004 Guichon and Thomas~\cite{Guichon:2004xg} showed that by
expanding about $<\sigma>=0$ one could derive an effective force
of the Skyrme type (widely used in nuclear structure calculations)
in which the local two-body effective interaction was supplemented
by three- and four-body forces, proportional to $d$ and $d^{2}$,
respectively. A comparison between the various coefficients in the
SkIII force with those derived from the QMC model showed agreement
at typically the 10\% level. This remarkable result provided the first,
direct microscopic connection between a commonly used effective nuclear
force and the underlying degrees of freedom of QCD.

As important as the results of Ref.~\cite{Guichon:2004xg} were,
many important issues were left unaddressed: 

\begin{itemize}
\item The relatively large scalar field found in nuclear matter means that
one would really like to remove the need to expand about $\sigma=0$. 
\item Modern Skyrme forces tend to include density dependent coefficients
(fit to appropriate nuclear data), rather than being density independent
with an explicit three-body term. Hence one would really like to rewrite
the energy functional for QMC in terms of a density dependent effective
force. 
\item A crucial issue at higher densities, when for example one is calculating
the equation of state (EoS) for application to neutron stars, concerns
the density at which hyperons appear in matter in $\beta$-equilibrium.
For this purpose one would very much like to derive effective, density
dependent effective hyperon-nucleon and hyperon-hyperon forces. Because
one starts from the quark level, QMC is ideally suited for this purpose. 
\end{itemize}

\begin{figure}
\begin{center}\includegraphics[%
  clip,
  width=13cm,
  keepaspectratio]{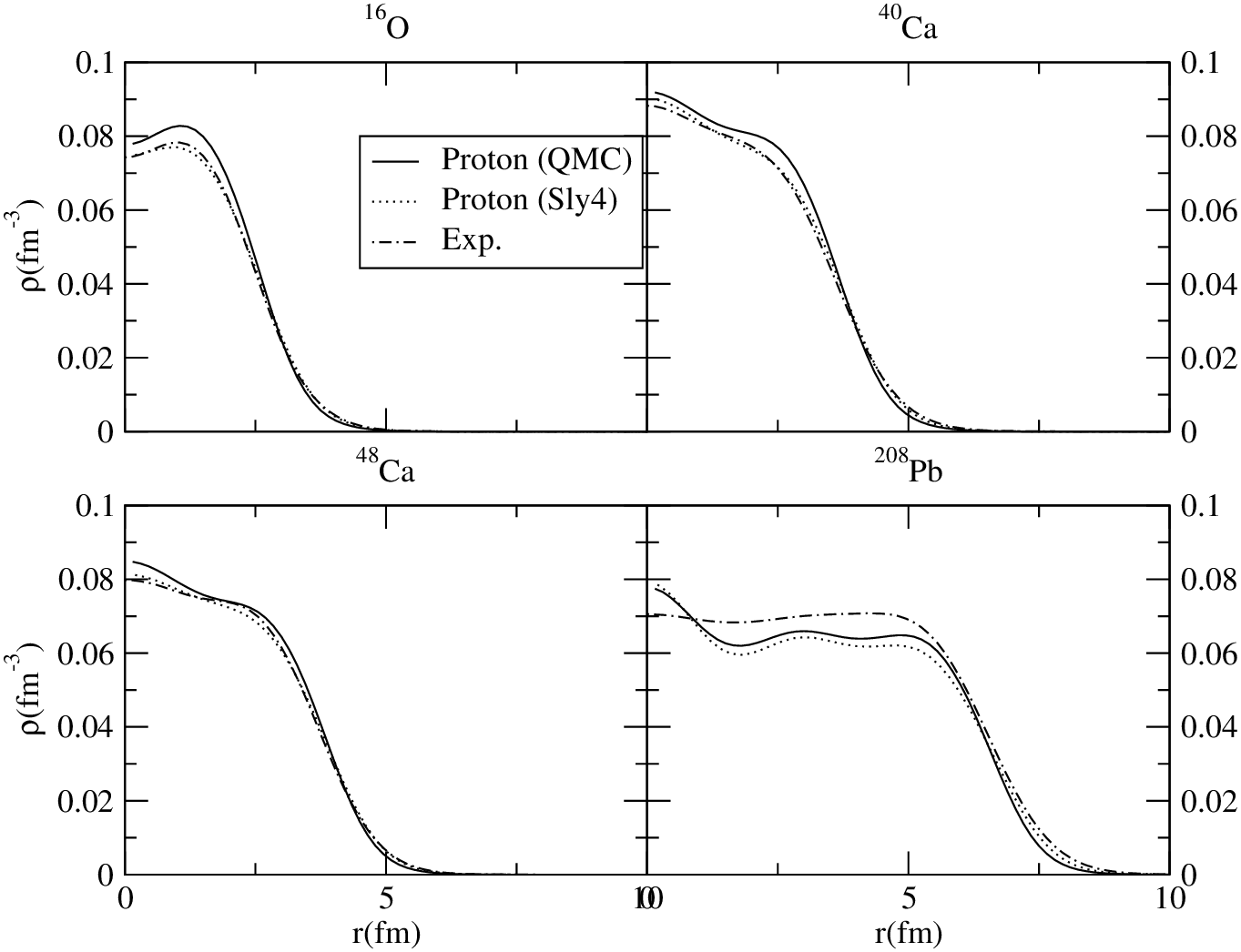}\end{center}

\caption{\label{fig:PDensity}Proton densities calculated using the density
dependent effective interaction derived from the QMC model compared
with experiment and with the predictions of the Skyrme Sly4 force -- from
Ref.~\cite{Guichon:2006er} .}
\end{figure}

All of these issues were successfully resolved by the recent formulation
of Guichon \textit{et al.}~\cite{Guichon:2006er}. The
resulting energy functional in its full relativistic form was too
difficult to use in finite nuclei. However, in that case a non relativistic
expansion was allowed and that provided an effective force with a novel
density dependence -- the coefficients being rational functions of
the local density. For example, the central pieces of the corresponding
Hamiltonian \begin{equation}
<H(\vec{r})>=\rho M+\frac{\tau}{2M}+{\mathcal{H}}_{0}+{\mathcal{H}}_{3}+{\mathcal{H}}_{eff}+{\mathcal{H}}_{fin}+{\mathcal{H}}_{SO}\label{eq:EffHam}\end{equation}
 took the form: \begin{eqnarray}
{\mathcal{H}}_{0}+{\mathcal{H}}_{3} & = & {\rho}^{2}\,\left[\frac{-3\,{G_{\rho}}}{32}+\frac{{G_{\sigma}}}{8\,{\left(1+d\,\rho\,{G_{\sigma}}\right)}^{3}}-\frac{{G_{\sigma}}}{2\,\left(1+d\,\rho\,{G_{\sigma}}\right)}+\frac{3\,{G_{\omega}}}{8}\right]+\nonumber \\
 &  & {\left({{\rho}_{n}}-{{\rho}_{p}}\right)}^{2}\left[\frac{5\,{G_{\rho}}}{32}+\frac{{G_{\sigma}}}{8\,{\left(1+d\,\rho\,{G_{\sigma}}\right)}^{3}}-\frac{{G_{\omega}}}{8}\right]\,.\label{eq:QMC_dd}\end{eqnarray}
 This appears rather different from commonly used density dependent
Skyrme forces such as SkM, which includes a dependence on a fractional
power of the density. For comparison, we show for SkM the 
central pieces corresponding to those that were given for QMC 
in Eq.~(\ref{eq:QMC_dd}): \be
{\mathcal{H}}_{0}+\mathcal{H}_{3} = \frac{{\rho}^{\frac{1}{6}}\,{t_{3}}\,
\left(2\,
{\rho}^{2}-{{\rho}_{n}}^{2}-{{\rho}_{p}}^{2}\right)}{24}+\frac{{t_{0}}\,
\left(
{\rho}^{2}\,\left(2+{x_{0}}\right)-\left(1+2\,{x_{0}}\right)\,
\left({{\rho}_{n}}^{2}
+{{\rho}_{p}}^{2}\right)\right)}{4} \, . \label{eq:SkM} \ee
In order to make a quantitative comparison between these two forms
of effective potential the SkM form was fit to the QMC form over the
density range $(0,\rho_{0})$. The comparison between the usual phenomenological
values of the constants appearing in the SkM force and those fitted
to QMC showed a level of agreement at the 10-20\% level. This is reasonable
but by no means impressive. On the other hand, since the functional
forms are quite different it seemed appropriate to compare the predictions
of the two effective forces for real nuclear properties. Accordingly
the effective force derived from the QMC model was employed within
the Hartree-Fock framework to calculate the properties of closed shell,
finite nuclei. The result was extremely gratifying, with the level
of agreement between theory and data (where available and phenomenological
Skyrme-type forces where data was unavailable) 
being very good. In Fig.~\ref{fig:PDensity}
we show a comparison between the charge densities of selected closed
shell nuclei calculated within QMC in comparison with experimental
data, as well as the widely used Sly4 force~\cite{Chabanat:1997un}.
Clearly the level of agreement between the two theories is remarkably
good. Given that the QMC calculation had no parameters adjusted to
reproduce the properties of finite nuclei, the agreement with  
the experimental charge densities is also very satisfactory. 

\begin{figure}
\begin{center}\includegraphics[%
  clip,
  width=12cm,
  keepaspectratio]{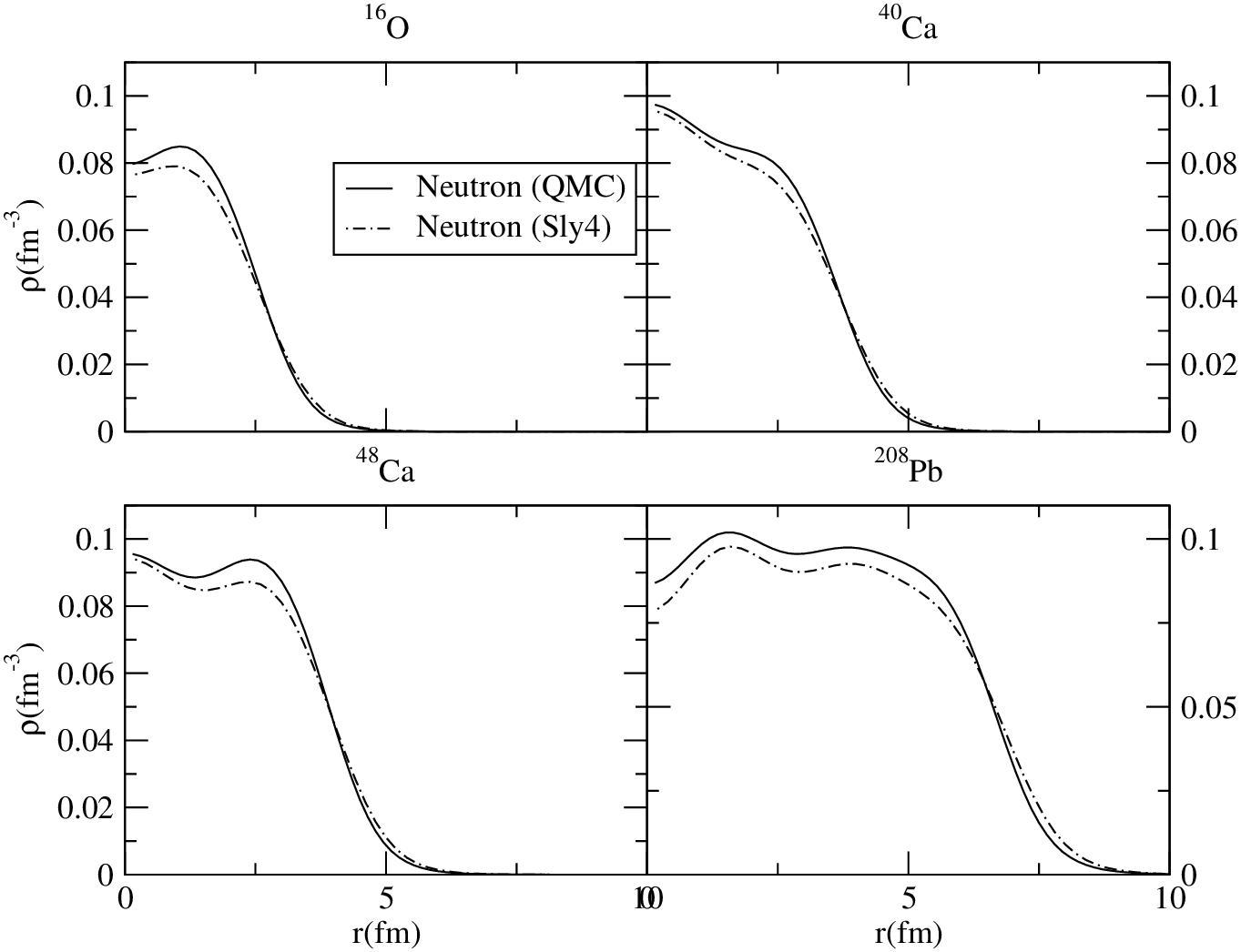}\end{center}

\caption{\label{fig:NDensity}Neutron densities calculated using the density
dependent effective interaction derived from the QMC model compared
with the predictions of the Skyrme Sly4 force -- from Ref.~\cite{Guichon:2006er}.}
\end{figure}

As there is no model independent experimental
determination of the neutron densities we show, in Fig.~\ref{fig:NDensity},
a comparison between the neutron densities produced by Sly4 and by
the QMC model. Once again, it is reassuring in terms of assessing
the quality of the effective interaction derived from the QMC model
that they are so close. 

Another key test of the model is its ability to describe the dependence
of the spin-orbit splittings on both mass number and isospin. At first
glance this looks like an area in which QMC might fail, in view
of the strong isospin dependence of the effective force derived from
QMC. However, as we see in Table~\ref{tab:Splittings} (taken from
Ref.~\cite{Guichon:2006er}), this worry is in fact misplaced. Once
the interaction is used, within the Hartree-Fock framework, to self-consistently
determine the nuclear structure there is agreement between the predictions
of QMC and experiment across the full range of nuclear mass number
for both protons and neutrons. 
\begin{table}
\begin{center}\begin{tabular}{|c|c|c|c|c|}
\hline 
&
 Neutrons (exp)&
 Neutrons (th)&
 Protons (exp)&
 Protons (th) \tabularnewline
\hline
$^{16}O,\,1p_{1/2}-1p_{3/2}$&
 6.10&
 6.01&
 6.3&
\tabularnewline
\hline
$^{40}Ca,\,1d_{3/2}-1d_{5/2}$&
 6.15&
 6.41&
 6.00&
 6.24 \tabularnewline
\hline
$^{48}Ca,\,1d_{3/2}-1d_{5/2}$&
 6.05 (Sly4)&
 5.64&
 6.06 (Sly4)&
 5.59\tabularnewline
\hline
$^{208}Pb,\,2d_{3/2}-2d_{5/2}$&
 2.15 (Sly4)&
 2.04&
 1.87 (Sly4)&
 1.74 \tabularnewline
\hline
\end{tabular}\end{center}

\caption{\label{tab:Splittings}Values of the spin-orbit splitting for selected
nuclear levels calculated from the QMC model, in comparison with the
corresponding experimental values, where known. As they are not so
well known in the case of $^{48}Ca$ and $^{208}Pb$, there we give
the values corresponding to the Skyrme Sly4 force. }
\end{table}

\section{Concluding Remarks}

\begin{figure}
\begin{center}\includegraphics[%
  clip,
  width=8cm,
  keepaspectratio]{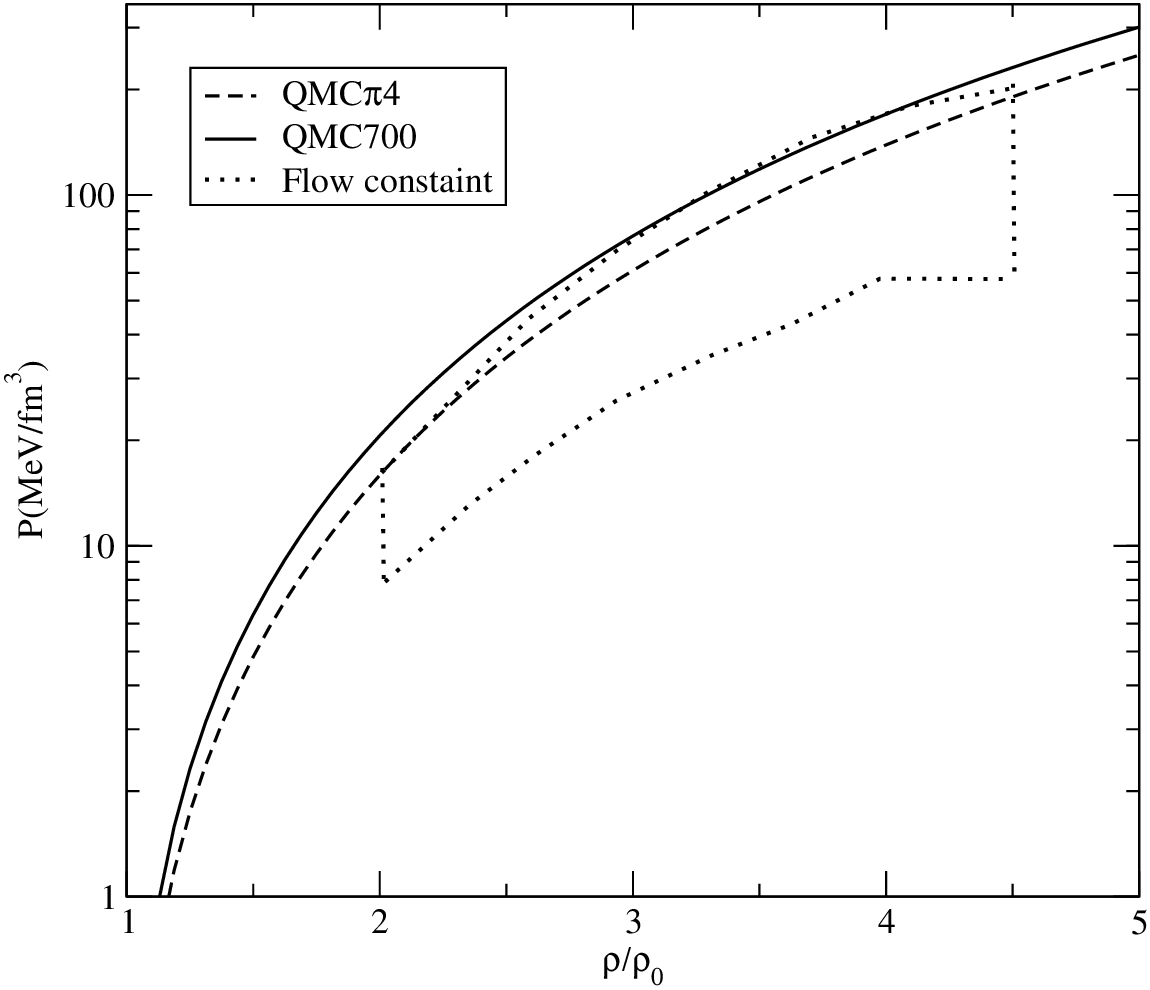}\end{center}

\caption{\label{cap:flow}QMC predictions for the pressure as a function 
of density in comparison with the allowed region for isoscalar matter 
deduced from the constraints of heavy-ion flow data~\cite{Daniel2002}. }
\end{figure}

In the limited space available we have been able to introduce
just the main ideas of these modern QMC based calculations. The results
obtained with the density dependent effective force are particularly
impressive. However, they are based upon the non-relativistic approximation,
which cannot be expected to be reliable much beyond $\rho_{0}$. Indeed,
as shown by Guichon \textit{et al.}~\cite{Guichon:2006er}, it leads
to an error of almost 50\% in the velocity of sound in the nuclear
medium at just $2\rho_{0}$. Thus, in studying dense
matter, one must return to using the original relativistic Hartree-Fock
formulation. This is feasible in uniform matter \cite{ToAppear}.
As an illustration of our results, in Figure \ref{cap:flow} we show 
our predictions for the pressure of symmetric nuclear matter as a
function of baryonic density. The labels QMC700 and QMC$\pi4$
refer to two extreme versions of the model which give 
an incompressibility of 340 MeV and 256 MeV, respectively. Also
shown in the figure is the experimental constraint deduced from the
flow data in high energy nuclear collisions~\cite{Daniel2002}.

Another challenge faced by nuclear theory, as the density increases, 
is the possible entry of hyperons into the nuclear stew. Hyperon-nucleon
forces are unfortunately poorly constrained by data, even in the case
of $\Lambda\, N$, while in the case of $\Sigma\, N$ and $\Xi\, N$
the situation is worse. Typical phenomenological forces rely on SU(3)
symmetry, a hard path to tread when cancellations between various
components of the force are so important. When it comes to using hypernuclear
data to constrain the parameters of Skyrme-type forces we note that
only a single $\Sigma$-hypernucleus has been confirmed and no $\Xi$-hypernuclei.
For $\Lambda$-hypernuclei the situation is a little better. It is
little wonder that there is no consensus on the threshold density
at which hyperons appear, nor even which one will appear first. As
the appearance of hyperons will soften the nuclear equation of state
and, for example, raise the critical density for the transition to quark
matter, this is a vital issue.

\begin{figure}
\begin{center}\includegraphics[%
  clip,
  width=15cm,
  keepaspectratio]{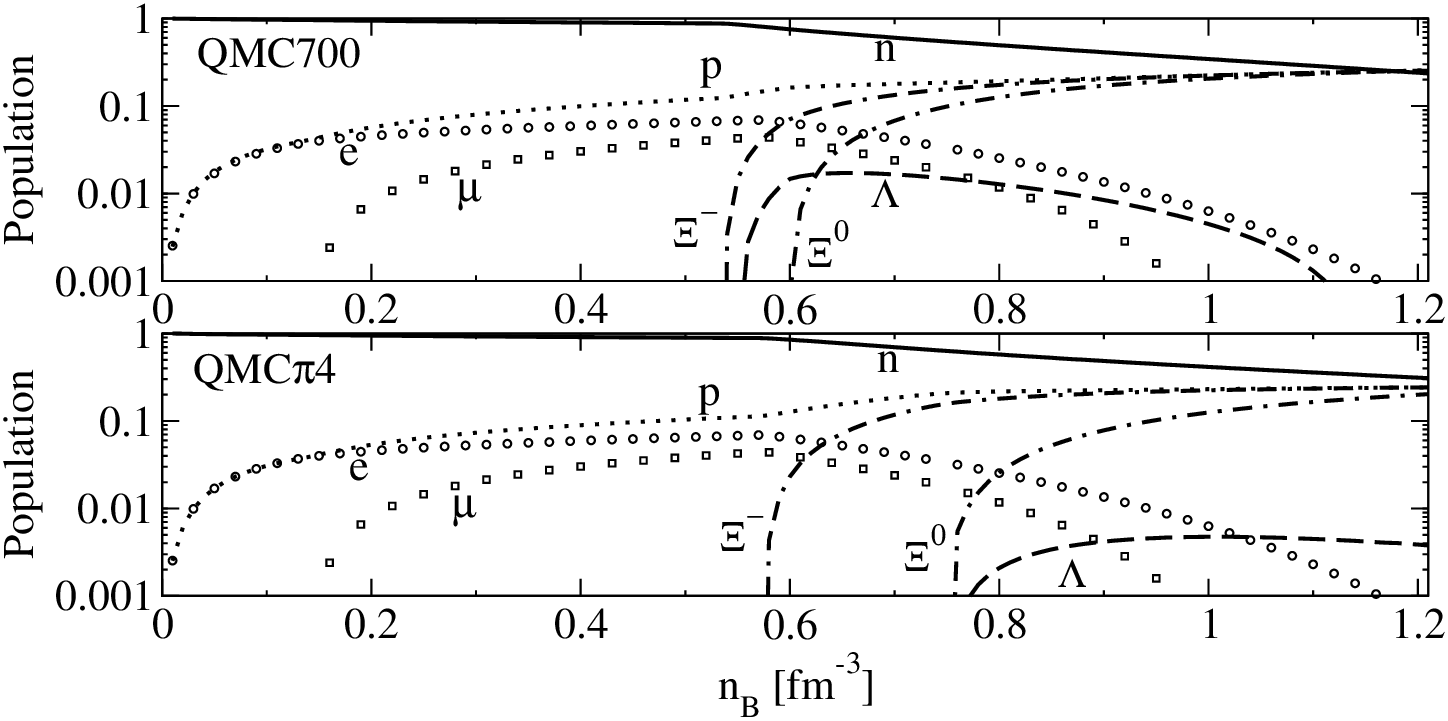}\end{center}

\caption{\label{cap:equilib} The $\beta$ equilibrium population
of particles calculated for the models QMC700 and QMC$\pi4$ 
-- from Ref.~\cite{ToAppear}.}
\end{figure}

Because the QMC model starts at the quark level,
it can can also be used to calculate density dependent, effective
hyperon-nucleon or even hyperon-hyperon forces -- with
no additional parameters. The only hypothesis needed is
that the nuclear fields do not couple to the strange quark. This
is certainly a good approximation if the meson fields simulate 
multi-pion exchanges. It is also supported by the absence of spin-orbit
splitting in $\Lambda$ hypernuclei. As shown in Figure~\ref{cap:equilib},
a striking prediction of the model is that the thresholds for the
appearance of strange particles do not follow the usual pattern
where the $\Sigma^{-}$ and $\Lambda$ appear first. Because of the scalar
polarizability of the baryons and the use of the Hartree-Fock approximation
\cite{ToAppear} (rather than the mean field approximation), the $\Xi$ and $\Lambda$
appear first and at a rather high density ($\sim 3\rho_{0})$. The
$\Sigma$ does not show up in the density domain pertinent to neutron
stars. The resulting equation of state is relatively stiff and produces 
a maximum neutron star mass of about $2.1M_{\odot}$, with a rather
small central density ($\sim6\rho_{0})$. 

Finally, we note that the QMC model is built upon the MIT bag model.
This is a fairly ancient and unsophisticated quark model and although
all indications are that the details of the model are not so important
one would like to do better. In particular, one would ideally like
to have a quark model describing the structure of the hadrons which
is covariant, in order to address the problem of nuclear structure
functions -- the famous EMC effect~\cite{Aubert:1983xm,Geesaman:1995yd}
-- with some confidence.

In parallel with the developments concerning QMC, Bentz, Thomas and
collaborators have made considerable progress with just such a model.
They took the NJL model, modified using proper time regularization
in order to simulate confinement. The model is therefore covariant,
confining and respects chiral symmetry. Exactly as in QMC, the hadron
structure in-medium was self-consistently modified (by solving the
corresponding Faddeev equations). Also as in the QMC model the scalar
polarizability led to the saturation of nuclear matter and, incidentally,
solved the long-standing problem of chiral collapse suffered by the
NJL model~\cite{Bentz:2001vc}. In application to the nuclear EMC
effect the model is able to reproduce the data for unpolarized structure
functions~\cite{Cloet:2005rt,Cloet:2006bq}. However, even more important, 
it makes a prediction characteristic of models of the QMC type in
which the mean scalar field enhances the lower Dirac components of
the confined quark wave functions. (We note that a similar calculation
within the chiral quark soliton model~\cite{Smith:2005ra} produces
a qualitatively similar result.) That is, it predicts a polarized
EMC effect roughly twice as big as the unpolarized effect. This will
provide a critical test of such models.

A final success of this approach has been to use the same underlying
quark model to describe matter made of {}``nucleons'' (albeit with
modified internal structure), as well as matter 
made of quarks~\cite{Lawley:2006ps,Lawley:2005ru}.
Indeed a major discovery of Lawley and collaborators~\cite{Lawley:2006ps}
was that, once the effect of the pion cloud of the nucleon~\cite{Young:2002cj}
is taken into account, one can use the same underlying, quark level
Hamiltonian to consistently describe the hadronic, quark matter and
superconducting quark matter phases of dense matter. This is a major
step forward in a field where one is usually forced to employ completely
different models for the different phases. Such calculations are of
great interest in view of the difficulties in simulating supernova
explosions, as well as because of the very different mass-radius relations
that one typically finds for stars containing quark matter.

\section*{Acknowledgements}

It is a pleasure to acknowledge the collaboration with W.~Bentz,
I. Clo\"et, S.~Lawley, H.~Matevosyan, J.~Riskova Stone, 
K.~Saito, N.~Sandulescu and K.~Tsushima,
who made crucial contributions to the work described here. This work
was supported by DOE contract DE-AC05-06OR23177, under which Jefferson
Science Associates operates Jefferson Lab.


\begin{thebibliography}{10}
\bibitem{Cottingham:1973wt}W.~N.~Cottingham, M.~Lacombe, B.~Loiseau, J.~M.~Richard and
R.~Vinh Mau, Phys.\ Rev.\ D \textbf{8} (1973) 800. 
\bibitem{Machleidt:1987hj}R.~Machleidt, K.~Holinde and C.~Elster, Phys.\ Rept.\ {} \textbf{149}
(1987) 1. 
\bibitem{deSwart:1995mb}J.~J.~de Swart, R.~A.~M.~Klomp, M.~C.~M.~Rentmeester and T.~A.~Rijken,
Few Body Syst.\ Suppl.\ {} \textbf{8} (1995) 438 {[}arXiv:nucl-th/9509024{]}. 
\bibitem{Wiringa:1994wb}R.~B.~Wiringa, V.~G.~J.~Stoks and R.~Schiavilla, Phys.\ Rev.\ C
\textbf{51} (1995) 38 {[}arXiv:nucl-th/9408016{]}. 
\bibitem{Carlson:1983kq}J.~Carlson, V.~R.~Pandharipande and R.~B.~Wiringa, Nucl.\ Phys.\ A
\textbf{401} (1983) 59. 
\bibitem{Epelbaum:2004fk}E.~Epelbaum, W.~Glockle and U.~G.~Meissner, Nucl.\ Phys.\ A
\textbf{747} (2005) 362 {[}arXiv:nucl-th/0405048{]}; 
\bibitem{Epelbaum:2005pn}E.~Epelbaum, Prog.\ Part.\ Nucl.\ Phys.\ {} \textbf{57} (2006)
654 {[}arXiv:nucl-th/0509032{]}. 
\bibitem{Serot:1984ey}B.~D.~Serot and J.~D.~Walecka, Adv.\ Nucl.\ Phys.\ {} \textbf{16}
(1986) 1. 
\bibitem{Brockmann:1996xy}R.~Brockmann and R.~Machleidt, arXiv:nucl-th/9612004. 
%
\bibitem{Guichon:1995pu}
  P.~A.~M.~Guichon, G.~Q.~Liu and A.~W.~Thomas,
  Nucl.\ Phys.\ A {\bf 591}, 606 (1995)
  [arXiv:nucl-th/9605031].
%
\bibitem{Thomas_Weise}A.~W.~Thomas and W.~Weise, {}``The Structure of the Nucleon'',
VCH Verlag, Berlin (2001)
\bibitem{Guichon:1987jp}P.~A.~M.~Guichon, Phys.\ Lett.\ B \textbf{200} (1988) 235. 
\bibitem{Guichon:1995ue}P.~A.~M.~Guichon, K.~Saito, E.~N.~Rodionov and A.~W.~Thomas,
Nucl.\ Phys.\ A \textbf{601} (1996) 349 {[}arXiv:nucl-th/9509034{]}. 
\bibitem{Saito:2005rv}K.~Saito, K.~Tsushima and A.~W.~Thomas, {}``Nucleon and hadron
structure changes in the nuclear medium and impact on observables,''
to appear in Prog.\ Part.\ Nucl.\ Phys.\  arXiv:hep-ph/0506314;
K.~Saito and A.~W.~Thomas, Phys.\ Rev.\ C \textbf{52} (1995)
2789 {[}arXiv:nucl-th/9506003{]}. 
%
\bibitem{Saito:1996yb}
  K.~Saito, K.~Tsushima and A.~W.~Thomas,
  Phys.\ Rev.\ C {\bf 55}, 2637 (1997)
  [arXiv:nucl-th/9612001].
%
\bibitem{Tsushima:1998qw}
  K.~Tsushima, D.~H.~Lu, A.~W.~Thomas and K.~Saito,
  Phys.\ Lett.\ B {\bf 443}, 26 (1998)
  [arXiv:nucl-th/9806043].
%
\bibitem{Tsushima:1997cu}
  K.~Tsushima, K.~Saito, J.~Haidenbauer and A.~W.~Thomas,
  Nucl.\ Phys.\ A {\bf 630}, 691 (1998)
  [arXiv:nucl-th/9707022].
%
\bibitem{Lu:1997mu}
  D.~H.~Lu, A.~W.~Thomas, K.~Tsushima, A.~G.~Williams and K.~Saito,
  Phys.\ Lett.\ B {\bf 417}, 217 (1998)
  [arXiv:nucl-th/9706043].
%
\bibitem{Saito:1993yw}K.~Saito and A.~W.~Thomas, Nucl.\ Phys.\ A \textbf{574} (1994)
659. 
\bibitem{Vautherin:1971aw}D.~Vautherin and D.~M.~Brink, Phys.\ Rev.\ C \textbf{5} (1972)
626. 
\bibitem{Guichon:2004xg}P.~A.~M.~Guichon and A.~W.~Thomas, Phys.\ Rev.\ Lett.\ {}
\textbf{93} (2004) 132502 {[}arXiv:nucl-th/0402064{]}. 
\bibitem{Guichon:2006er}P.~A.~M.~Guichon, H.~H.~Matevosyan, N.~Sandulescu and A.~W.~Thomas,
Nucl.\ Phys.\ A \textbf{772} (2006) 1 {[}arXiv:nucl-th/0603044{]}. 
\bibitem{Chabanat:1997un}E.~Chabanat, P.~Bonche, P.~Haensel, J.~Meyer and R.~Schaeffer,
Nucl.\ Phys.\ A \textbf{635} (1998) 231. 
\bibitem{ToAppear}P.~A.~M.~Guichon, H.~Matevosyan, J.~Riskova Stone and A.~W.~Thomas,
to be published, {[}arXiv:nucl-th/0611030{]}.
\bibitem{Daniel2002}P. Danielewicz, R. Lacey and W.G. Lynch, Science 298 (2002) 1592.
\bibitem{Aubert:1983xm}J.~J.~Aubert \textit{et al.} {[}European Muon Collaboration{]},
Phys.\ Lett.\ B \textbf{123} (1983) 275. 
\bibitem{Geesaman:1995yd}D.~F.~Geesaman, K.~Saito and A.~W.~Thomas, Ann.\ Rev.\ Nucl.\ Part.\ Sci.\ {}
\textbf{45} (1995) 337. 
\bibitem{Bentz:2001vc}W.~Bentz and A.~W.~Thomas, Nucl.\ Phys.\ A \textbf{696} (2001)
138 {[}arXiv:nucl-th/0105022{]}. 
\bibitem{Cloet:2005rt}I.~C.~Cloet, W.~Bentz and A.~W.~Thomas, Phys.\ Rev.\ Lett.\ {}
\textbf{95} (2005) 052302 {[}arXiv:nucl-th/0504019{]}. 
\bibitem{Cloet:2006bq}I.~C.~Cloet, W.~Bentz and A.~W.~Thomas, arXiv:nucl-th/0605061. 
\bibitem{Smith:2005ra}J.~R.~Smith and G.~A.~Miller, Phys.\ Rev.\ C \textbf{72} (2005)
022203 {[}arXiv:nucl-th/0505048{]}. 
\bibitem{Lawley:2005ru}S.~Lawley, W.~Bentz and A.~W.~Thomas, Phys.\ Lett.\ B \textbf{632}
(2006) 495 {[}arXiv:nucl-th/0504020{]}. 
\bibitem{Lawley:2006ps}S.~Lawley, W.~Bentz and A.~W.~Thomas, J.\ Phys.\ G \textbf{32}
(2006) 667 {[}arXiv:nucl-th/0602014{]}. 
\bibitem{Young:2002cj}R.~D.~Young, D.~B.~Leinweber, A.~W.~Thomas and S.~V.~Wright,
Phys.\ Rev.\ D \textbf{66} (2002) 094507 {[}arXiv:hep-lat/0205017{]}. 
\end{thebibliography}
\end{document}